\documentclass[aps,twocolumn,amsmath,amssymb,nofootinbib,superscriptaddress]{revtex4-2}
\usepackage{adjustbox}
\usepackage[utf8]{inputenc} % allow utf-8 input
\usepackage[colorlinks=true,
					linkcolor=magenta,
					anchorcolor=magenta,
					citecolor=magenta,
					urlcolor=magenta]{hyperref}        % hyperlinks
\usepackage{url}            % simple URL typesetting
\usepackage{booktabs}       % professional-quality tables
\usepackage{amsfonts}       % blackboard math symbols
\usepackage{nicefrac}       % compact symbols for 1/2, etc.
\usepackage{microtype}      % microtypography
\usepackage{natbib}

\usepackage{bm}
\usepackage{bbm}
\usepackage{braket}
\usepackage{graphicx}
\usepackage{subfig}
\usepackage{float}
\usepackage{amsthm}
\usepackage{algorithmic}
\usepackage[linesnumbered,ruled]{algorithm2e}
\SetKwInOut{Parameter}{parameter}

\usepackage{mathtools}
\usepackage{nccmath}
\usepackage{color}
\usepackage{caption}
\usepackage{titlesec}

\usepackage{soul}
\titlespacing{\title}{0pc}{0.1pc}{0.3pc}
\titlespacing{\section}{0pc}{0.1pc}{0.3pc}

\begin{document}

\title{Transmission and generation of arbitrary W states via an optomechanical interface}

\author{Rui-Xia Wang}
\email{wangrx.2009@tsinghua.org.cn}

\affiliation{Beijing Academy of Quantum Information Sciences, Beijing,100193, China}

\affiliation{School of Natural Sciences, University of California, Merced, California 95343, USA}

\date{\today}

\pacs{xxx}

\begin{abstract}

We propose a universal and nontrivial scheme to transmit and generate an arbitrary W state for multiple cavities via an optomechanical interface. In transmission and generation processes, high fidelity can
be obtained by optimizing the time-dependent coupling strengths between
the cavities and the mechanical resonator. With a group of optimal couplings, an arbitrary entangled W state in the multipartite system can be mapped to the pulse shape of a single photon and transmitted out of the system. In the time reversal process, an arbitrary W state can be generated with an incident single photon with certain pulse shape. The functions of the optimal couplings, which are used for both transmission and generation processes, only depend on the parameters of the system, which does not change with the arbitrary entangled W states and the pulse shape of the single photons.

\end{abstract}

\maketitle

\section{Introduction}

Quantum entanglement involves non-local correlations between subsystems,
which has been recognized as one of the core resources in quantum
technologies \cite{horodecki2009quantum,brunner2014bell,gisin2002quantum,ladd2010quantum}. The bipartite entanglement state has been studied extensively, however, multipartite entanglement has more complicated structure and has been experimentally prepared in various quantum systems \cite{monz201114,omran2019generation,song2019generation}. As is well known, the GHZ states and W states are two representative multipartite entangled states. A remarkable property of the W state is its robustness against losses of qubits, since tracing out any part from a W state, there exists entanglement in the rest parties \cite{dur2000three,koashi2000entangled}. This property makes W state a useful entanglement resources in quantum information\cite{joo2003quantum,agrawal2006perfect}. To date a number of theoretical schemes have been proposed to generate W states in various systems \cite{bastin2009operational,kang2016fast,ccakmak2019robust,sharma2020generation,kim2020efficient,blasiak2021efficient,shao2009one,li2018dissipation}, and experimentally, there are a few implementations for W states in superconducting qubits \cite{neeley2010generation}, photons \cite{papp2009characterization}, trapped ions \cite{haffner2005scalable} and atomic ensembles \cite{choi2010entanglement}. The entangled W state can also be generated via optomechanical interface.

Optomechanical interactions can take place via the radiation pressure force induced by the optical fields. The optomechanical interface, in which the mechanical oscillator couples to the optical cavities or microwave cavities, can mediate a quantum state transfer between light and matter or microwave and optical fields \cite{dong2012optomechanical,wang2012using,tian2012adiabatic,bochmann2013nanomechanical,palomaki2013coherent,andrews2014bidirectional,dong2015optomechanical}. The entanglement generation via optomechanical interface has also been studied intensively between two cavity modes\cite{barzanjeh2011entangling,barzanjeh2019stationary}, one cavity mode and one mechanical mode\cite{hofer2011quantum,tian2013robust,palomaki2013entangling,wang2014nonlinear,yang2017generation} or two cavity modes and one mechanical mode \cite{wang2013reservoir,wang2015bipartite}. It is still appealing to propose a method to map the multipartite entangled states prepared in multiple microwave or optical cavities to the pulse shape of single photons via the optomechanical interface.

An arbitrary multipartite entangled W state for the cavity modes $a_i$ ($i=1,2,...,n$) can be expressed as $W_n=w_1|100...0\rangle_{a_1a_2...a_n}+w_2|010...0\rangle_{a_1a_2...a_n}+...+w_n|000...1\rangle_{a_1a_2...a_n}$ ($\sum_{i=1}^{n}|w_i|^2=1$), which may be realized in the experiment with the microchip device\cite{barzanjeh2019stationary,pinske2020highly} or by applying the piezo-optomechanical structure\cite{balram2016coherent,mirhosseini2020superconducting}. In this work, we propose a method to transmit and generate an arbitrary entangled W state in an open multipartite system composed of multiple cavities coupling to an optomechanical interface. The whole process for our scheme is as follows: 1. There is an arbitrary W state in system A as shown in Fig. \ref{f1}(a), and system A can be part of a larger quantum system. 2. We can map the arbitrary W state to the pulse shape of a single photon leaving from system A with a set of optimized time-dependent couplings. 3. In system B, we can produce the W state which is the same as the previous one in system A by applying a set of time-dependent couplings which are time reversal of the output process. In this way, an entangled arbitrary W state can be transmitted between two distant quantum systems with only one output-and-input process. In this proposal, the cavities used for the optomechanical interface or preparing the W state can work at optical or microwave frequencies, and the quantum states can be converted between single photons and cavity modes with vastly different frequencies. In our scheme, we suppose that, the optomechanical interface is composed of a mechanical resonator and an optical cavity, and the cavities for the W states preparation are microwave cavities. In the transmission and generation processes, the strength of the couplings between the cavities and the mechanical resonator are time-dependent and optimized. With a group of optimal time-dependent couplings, the mapping and transmission of an arbitrary entangled W state can be proceeded, and the time reversal process is just the generation process for the multipartite entangled W state. This method can be used to transfer quantum information between different quantum systems. In Section \ref{s2} of this article, we describe the basic model of a multi-cavity system. The linearized Hamiltonian and the dressed states for this system are given. In Section \ref{3a}, we give the main theoretical results for the time-independent process. In time-independent case, only when the system is in a certain initial state, the probability of transmitting the W state out of the system will be the unit, and if not, the probability will be smaller than $1$. In Section \ref{3b}, we give the theoretical method of transmitting an arbitrary W state from one system to another with time-dependent couplings. In Section \ref{s4}, we build a model to calculate the fidelity of the input process with a full quantum process. In this model, we assume that the mechanical resonator can be cooled to the ground state initially. In Section \ref{s5} and \ref{s6}, the transmission and generation process are simulated numerically, and finally, we talk about the fidelities of the generation process with and without system dampings. In Section \ref{s7}, we give the analysis of the impact from the mechanical noise and propose the methods to enhance the fidelity for the entangled state transmission and generation processes.

\bigskip

\section{Model}\label{s2}
\begin{figure}[h]
\centering{\includegraphics[width=85mm]{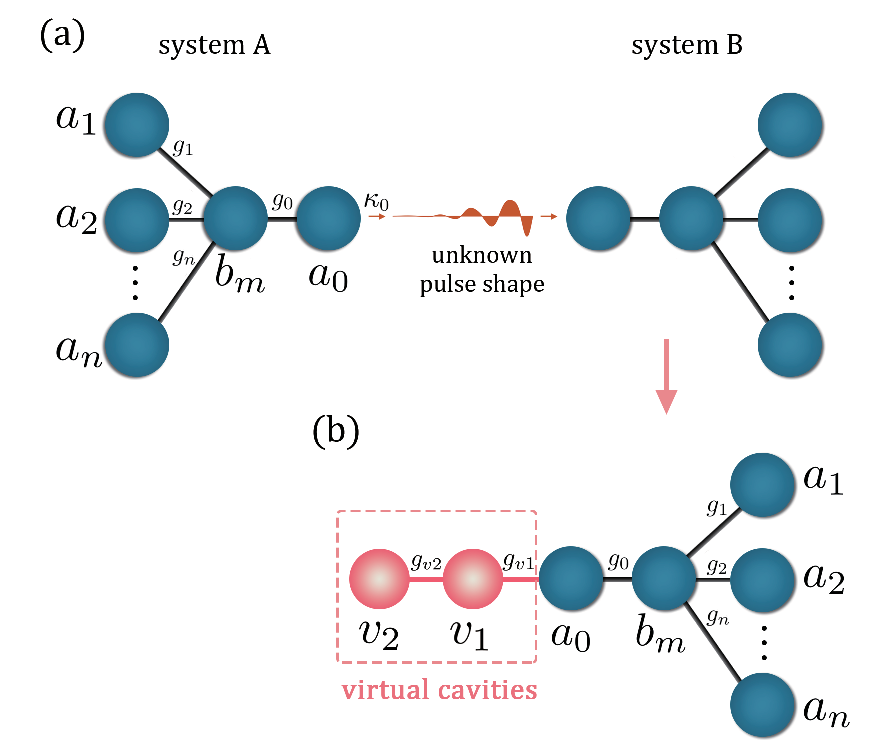}}
\caption{(a) The systems A and B are two hybrid systems, each of them is composed of $n$ microwave cavities coupling to an optomechanical interface. (b) Model used to calculate the generation process. We add two virtual cavities which coupled to the hybrid system to model the input single photon with arbitrary wave function. In this way, we can calculate the fidelity of the final state with a full quantum model.}
\label{f1}
\end{figure}

The coupling scheme we investigated is illustrated in Fig. \ref{f1}. The microwave cavities $a_i$ ($i=1,2,...,n$) with cavity frequency $\omega_i$ and optical cavity $a_0$ with frequency $\omega_c$ are coupled to a mechanical mode $b_m$ with mechanical frequency $\omega_m$. The coupling strength between the microwave cavity mode $a_i$ and the mechanical mode $b_m$ is $g_i$ with the driving pulse on the microwave cavity with the frequency $\omega_{d_i}$, and the coupling strength between the optical mode and the mechanical mode is $g_0$ with the driving pulse on the optical cavity with frequency $\omega_{d_0}$. The coupling strength can be enhanced by the driving pulse and be tuned by varying it. The mechanical resonator and optical cavity compose an optomechanical interface with the optical cavity damping rate $\kappa_0$ and mechanical damping rate $\gamma_m$. The damping rate of the microwave cavity $a_i$ is $\kappa_i$, and the condition $\kappa_0\gg\gamma_m,\kappa_i$ is saticfied to realize a high fidelity for the entangled state transmission and generation processes, as the value of $\kappa_0$ determines the efficiency of the single-photon output and input processes. And $\omega_m\gg\kappa_0$ is also satisfied to reach the resolved sideband limit \cite{teufel2008dynamical}. Initially, before the transmission process, the mechanical resonator can be cooled and optically damped via the driven cavity $a_0$, then we can assume that, the initial state of the mechanical resonator is in the ground state\cite{teufel2011sideband,chan2011laser,wang2013reservoir}. Assume that, the pump fields are red detuned as $\omega_j-\omega_{d_j}=\omega_m$ ($j=0,1,2,...,n$). And for simplicity, all of the values of the couplings are real, then the linearized interaction Hamiltonian for the closed system is ($\hbar=1$) \cite{aspelmeyer2014cavity, tian2015optoelectromechanical}

\begin{equation}
\hat{H}_{I}=\sum_{j=0}^{n}g_{j}\left(\hat{a}_{j}^{\dagger}\hat{b}_{m}+\hat{b}_{m}^{\dagger}\hat{a}_{j}\right),
\end{equation}
where $\hat{a}_j$ ($j=0,1,2,...,n$) and $\hat{b}_m$ are the annihilation operators of the cavity modes and mechanical mode, respectively. In the basis of $\left|100...0\right\rangle _{a_{0}a_{1}...a_{n}b_{m}}$,
$\left|010...0\right\rangle _{a_{0}a_{1}...a_{n}b_{m}}$,..., $\left|000...1\right\rangle _{a_{0}a_{1}...a_{n}b_{m}}$, this interaction Hamiltonian can be written as a matrix with $n+2$ eigenvalues, $n$ of them are degenerate with the eigenvalues of $\lambda_i=0$ ($i=1,2,...,n$), the other two eigenvalues are $\lambda_{n+1}=s_n$ and $\lambda_{n+2}=-s_n$, where $s_{n}=\sqrt{\sum_{j=0}^{n}g_{j}^{2}}$. All of the eigenvalues define $n+2$ adiabatic eigenstates. In the adiabatic basis, the Hilbert space is decomposed into two subspaces: the eigenstates $|\phi_i\rangle$ ($i=1,2,...,n$) with the eigenvalues of $\lambda_i=0$, these $n$ eigenstates are dark states, for which, the probability for the mode $b_m$ being excited is $0$. The dark states can be given by $|\phi_{1}\rangle=\frac{1}{s_{1}}[g_1,-g_0,0,0...]^{T}$, $|\phi_{2}\rangle=\frac{1}{s_1s_2}[g_0g_2,g_1g_2,-s_1^2,0,0,...]^{T}$, $|\phi_{3}\rangle=\frac{1}{s_2s_3}[g_0g_3,g_1g_3,g_2g_3,-s_2^{2},0,0...]^{T}$,..., $|\phi_{n}\rangle=\frac{1}{s_{n-1}s_n}[g_0g_n,g_1g_n,...,-s_{n-1}^2,0]^{T}$. Another subspace is a two-dimensional bright space, the two bright eigenstates can be expressed as $|\phi_{n+1}\rangle =\frac{1}{\sqrt{2}s_n}[g_0,g_1,g_2,...,g_n,s_n]^{T}$ and $|\phi_{n+2}\rangle =\frac{1}{\sqrt{2}s_n}[g_0,g_1,g_2,...,g_n,-s_n]^{T}$, where $s_{i}=\sqrt{\sum_{j=0}^{i}g_{j}^{2}}$ ($i=0,1,2,...,n$) \cite{kis2001nonadiabatic}.

\bigskip

\section{Adiabatic output process}\label{s3}

The dissipative quantum systems are usually described by a Lindblad master equation. In this equation, there are two dissipative terms, one is about the coherent nonunitary evolution, and the other describes quantum jumps\cite{plenio1998quantum,akram2013entangled,mirza2014single,flayac2017unconventional}. According to the quantum jump theory, if until time $t$, there is no photon emitted from the system to the environment, the time evolution of the system from the beginning to time $t$ is a coherent decay process and can be described by a non-Hermitian Hamiltonian, which can be called conditional Hamiltonian (see Supplement 1 for more details). For the output process, we would like to calculate the ensemble averaged pulse shape of the output single photon under certain time-dependent or time independent couplings. From the input-output theory, the pulse shape of the single photon from cavity $a_0$ is equal to $\langle\hat{a}_{0,out}(t)\rangle$, and $\hat{a}_{0,out}(t)=-\sqrt{\kappa_{0}}\hat{a}_{0}(t)$. In order to get $\langle\hat{a}_{0}(t)\rangle$, we can use the Heisenberg-Langevin equations with system Hamiltonian\cite{ventura2019robust} and cavity decays or the Schr{\"o}dinger equation with conditional Hamiltonian\cite{li2006preparing}. According to quantum jump theory, considerring the coupling of the system to the output with the damping rates of $\kappa_j$ ($j=0,1,2,...,n$) and $\gamma_m$, the conditional Hamiltonian is  ($\hbar=1$) \cite{chen2021quantum, li2006preparing}

\begin{equation}
\hat{{H}}_{c}=-\sum_{j=0}^{n}\frac{i\kappa_{j}}{2}\hat{a}_j^{\dagger}\hat{a}_j-\frac{i\gamma_{m}}{2}\hat{b}_{m}^{\dagger}\hat{b}_{m}+H_I.
\end{equation}
Under the condition of $g_i\gg\kappa_i$ ($i=1,2,...,n$), for simplicity, we first neglect $\kappa_i$ and only consider the damping rates $\kappa_0$ and $\gamma_m$. In the basis of $\left|\phi_{1}\left(t\right)\right\rangle $, $\left|\phi_{2}\left(t\right)\right\rangle $,...,$\left|\phi_{n+2}\left(t\right)\right\rangle $,
the conditional Hamiltonian is an $n+2$ by $n+2$ matrix $\hat{\tilde{H}}_{c}^{[(n+2)\times(n+2)]}$ (see Supplement 1 for more details). In the adiabatic process, if initially, the system is in dark state, there will not be excitations with the bright states, so that, the evolution of the system will only depend on the first $n$ rows
and $n$ columns of $\hat{\tilde{H}}_{c}^{[(n+2)\times(n+2)]}$. We then extract the first $n$ rows and $n$ columns as $\hat{\tilde{H}}_{c}^{\left(n\times n\right)}=-\frac{i\kappa_0}{2}M$, where $M=(1-\frac{g_{0}^{2}}{s_{n}^{2}})|\phi_{0}\rangle \langle \phi_{0}|$ and $|\phi_{0}\rangle =\frac{1}{\sqrt{1-\frac{g_0^2}{s_n^2}}}[\phi_1^{(1)},\phi_2^{(1)},...,\phi_n^{(1)}]^{T}$, $\phi_i^{(1)}$ ($i=1,2,...,n$) is the first value of $|\phi_{i}\rangle$.

First, we study the coherent decay process without the incident pulse into the system. In general case, assume that, the wave function of the system is $|\psi(t)\rangle =\sum_{i=1}^{n}c_{i}(t)|\phi_{i}(t)\rangle $, substitute the wave function into the Schr{\"o}dinger equation, we can get
$i\frac{d}{dt}|\psi(t)\rangle =-\frac{i\kappa_0}{2}M(t)|\psi(t)\rangle $.
In the adiabatic approximation, there is $|\frac{1}{g_{i}(t)}\frac{dg_{i}(t)}{dt}|\ll |g_i(t)|$,
so that, $|\dot{\phi_i}(t)\rangle $ is negligible,
then there is

\begin{equation}
i\frac{d}{dt}C(t)=-\frac{i\kappa_0}{2}M(t)C(t),
\label{dc}
\end{equation}
where $C(t)=[c_1(t),c_2(t),...,c_n(t)]^{T}$.

There are $n$ eigenstates for the matrix $M(t)$, one is $|\varphi_{1}(t)\rangle =|\phi_{0}(t)\rangle$, with the eigenvalue $1-\frac{g_{0}^{2}}{s_{n}^{2}}$. The other $n-1$ eigenstates from $|\varphi_{2}(t)\rangle$ to $|\varphi_{n}(t)\rangle$ are degenerate with the same eigenvalues of $0$. Without loss of generality, we can define the matrix $U(t)=[\varphi_1(t),\varphi_2(t),...,\varphi_n(t)]$. The matrix $M(t)$ can be diagonalized by the matrix $U(t)$
and $U^{\dagger}(t)$ as $U^{\dagger}(t)M(t)U(t)=\Lambda(t)$,
where $\Lambda(t)$ is a diagonal matrix, $\Lambda_{11}(t)=1-\frac{g_{0}^{2}(t)}{s_{n}^{2}(t)}$ and $\Lambda_{ij}=0$ ($i,j=1,2,...,n$ and $ij\neq11$).

\subsection{Time-independent output process}\label{3a}
If at any time $t$, the coupling strength satisfies $g_{j}(t)=G(t)g_{j}(0)$ ($j=0,1,2,...,n$ and $G(t)$ is an arbitrary function and $G(t)$ cannot be $0$ all the time ), there is $\frac{dM(t)}{dt}=0_{n\times n}$. And equation (\ref{dc}) can be transformed into $i\frac{d\alpha(t)}{dt}=-\frac{i\kappa_0}{2}\Lambda\alpha(t)$, where $\alpha(t)=U^{\dagger}C(t)$. Solving this function, we can get the expression of the wave function as $C(t)=U\exp(\int_{0}^{t}-\frac{\kappa_0}{2}\Lambda dt^{\prime})U^{\dagger}C(0)$. At time $t=+\infty$, there is $\exp(-\frac{\kappa_0}{2}\Lambda_{11}t)=0$, then we can get $C(+\infty)=UIU^{\dagger}C(0)-|\phi_0\rangle\langle\phi_0|C(0)$. Assume that, initially, $C(0)=\beta_{1}|\phi_{0}\rangle +\beta_{2}|\tilde{\phi}_{0}\rangle $,
$\langle\tilde{\phi}_{0}|\tilde{\phi}_{0}\rangle =1$ and $\langle\phi_{0}|\tilde{\phi}_{0}\rangle =0$, then there is $C(+\infty)=\beta_{2}|\tilde{\phi}_{0}\rangle$.

If $\beta_{2}=0$, the initial state of the system is $|\phi_{0}\rangle$, we have $C(+\infty)=0_{n\times 1}$ and if $\beta_{2}\neq0$, there is $C(+\infty)=\beta_{2}|\tilde{\phi}_{0}\rangle$, which means that, when only considering the damping rates $\kappa_0$ and $\gamma_m$, in the time-independent case ($\frac{dM(t)}{dt}=0_{n\times n}$) and adiabatic transmission process, only when $C(0)=|\phi_0\rangle$, the final state of the system is empty, if not, at time $t=+\infty$, the probability of transmitting the state $C(0)$
out of the system is $|\beta_{1}|^{2}$, and the probability for the system to be at the state of $|\tilde{\phi}_{0}\rangle $ is $|\beta_{2}|^{2}$ .

\subsection{Time-dependent output process}\label{3b}
In the case of $\frac{dM(t)}{dt}\neq0$, equation (\ref{dc}) can be transformed into

\begin{equation}
\frac{d\alpha(t)}{dt}=-[\frac{\kappa_0}{2}\Lambda(t)-V(t)]\alpha(t),
\label{dadt}
\end{equation}
where $V(t)=\frac{dU^{\dagger}(t)}{dt}U(t)$ and $\alpha(t)=U^{\dagger}(t)C(t)$. Although the Hamiltonian of the system changes slowly,
there are $\Lambda_{ii}=0$ ($i=2,3,...,n$), so that $V(t)$
is not negligible. There is $\alpha(t)=[\alpha_1(t),\alpha_2(t),...,\alpha_n(t)]^{T}$, where $\alpha_i=\varphi_{i}^{\dagger}(t)C(t)$ ($i=1,2,...,n$) is the projection of $C(t)$ to $\varphi_{i}(t)$ at time $t$. Then there is

\begin{equation}
\frac{d\alpha_{i}(t)}{dt}=-\frac{\kappa_0}{2}\Lambda_{ii}(t)\alpha_{i}(t)+\sum_{j=1}^{n}V_{ij}(t)\alpha_{j}(t).
\end{equation}

We can prove that $V_{ij}(t)+V_{ji}(t)=0$ ($V(t)$ is real), and when $\kappa_0=0$, there is $d[\sum_{i=1}^{n}\alpha_{i}^{2}(t)]/dt=0$ (see Supplement 1 for more details). So that, in equation (\ref{dadt}), the effect of the part $-\frac{\kappa_0}{2}\Lambda\left(t\right)$ is to output the quantum state via $\varphi_{1}\left(t\right)$ and
the effect of $V(t)$ is to redistribute the populations
in every state $\varphi_{i}\left(t\right)$. Then we can conclude that, in case of $\frac{dM(t)}{dt}\neq0$, the initial state $C(0)$ can not determine
whether the state of the system can be output totally out of the system from the cavity mode $a_0$. Then we will study the methods to output or prepare an arbitrary entangled W state.

If there are several different initial states $C_{1}\left(0\right)$, $C_{2}\left(0\right)$,...,$C_{n}\left(0\right)$, we have that, for $\forall i\in(1,2,...,n)$ there is $\frac{dC_{i}(t)}{dt}=-\frac{\kappa_0}{2}M(t)C_{i}(t)$. For the initial state of $C_{i}(0)$, there will be a final state $C_{i}(T)$. Then if the initial state is $C_{s}(0)=\sum_{i=1}^{n}p_{i}C_{i}(0)$, there is $C_{s}(T)=\sum_{i=1}^{n}p_{i}C_{i}(T)$, where $|p_i|^2$ is the probability that the system is in state $C_{i}(t)$ at time $t$. When the $n$ initial states of $C_{i}\left(0\right)$ are linearly independent, if we can find a group of optimal time-dependent couplings $g_0(t),g_1(t),...,g_n(t)$ to totally output any one of $C_{i}(0)$ within a limited time of $[0,T]$ from the cavity mode $a_0$, which means that for $\forall i\in(1,2,...,n)$ there is $C_{i}\left(T\right)=0_{n\times 1}$, then, for an arbitrary initial state $C_{s}\left(0\right)$, with this group of optimal time-dependent couplings, there is $C_{s}\left(T\right)=0_{n\times 1}$.

With the above method, we can output arbitrary entangled W states in the microwave cavities $a_i$ ($i=1,2,...,n$) from the cavity mode $\hat{a}_0^{A}$ (the superscript $A$ represents the systme A shown in Fig. \ref{f1}(a)) and mapped them into the pulse shapes of the single photons with a certain group of optimal time-dependent couplings. The function of the output single photon from system $A$ can be denoted as $\hat{a}_{0,out}^{A}(t) = -\sqrt{\kappa_0}\hat{a}_0^{A}(t)$, and we define the average operator as $f(t)=-\sqrt{\kappa_0}\langle\hat{a}_0^{A}(t)\rangle$. The time reverse of the output process is just the entangled W state generation process. In the generation process, the initial state of the system is empty, a single photon with the pulse function of $\hat{a}_{0,in}^{B}(t)=\hat{a}_{0,out}^{A}(T-t)$ is inputted into the system $B$ from cavity mode $\hat{a}_0^{B}$. The driving pulse for the cavities $a_0,a_1,...,a_n$ are $g_0(T-t),g_1(T-t),...,g_n(T-t)$, respectively. And finally, the system will be in an entangled W state, which is the same with the initial state of the output process.

In fact, there is a trivial method to realize the entangled W state transmission and generation. In the output process, apart from $g_0$ being always not zero, there is only one coupling strength of $g_i$ ($i=1,2,...,n$) being not zero for a time. The state of the corresponding microwave cavity will be transmitted out of the system and the other $n-1$ cavities are decoupled from the system in this period of time. In this way, the state of all the microwave cavities will be transmitted out of the system one by one and the system will be empty at the final time. With this method, first, the microwave cavities should be decoupled totally from the hybrid system; second, the output process are not continuous, the pulse shape of the single photon is divided into $n$ time-bins artificially, and third, the microwave cavities is not equal in time sequence in the evolution process. To overcome these problems, we can apply the optimization method to optimize all the couplings at the same time to realize the sync output for all the microwave cavities.

\bigskip

\section{Adiabatic generation process}\label{s4}

In the input process, a single photon will be transmitted into the system from the cavity $a_0$ and generate a W state. There have been already some models to deal with similar cases\cite{duan2003cavity,kiilerich2019input}. In this article, we apply a simple and suitable model for our system to calculate the state evolving with a full quantum process. For the convenience of calculation, we can model the process of the single photon transmitting into the quantum system with two virtual quantum modes coupling to our target system. In an actual experiment, the virtual cavities do not exist, and they are just fictionalized for ease of calculation. And by calculating the time-dependent coupling strengths related to the virtual system, we can realize a quantum process as a single photon input a target system. The schematic diagram is shown in 
Fig. \ref{f1} (b). $v_1$ and $v_2$ are two virtual cavities, the coupling strength between cavities $v_1$ and $a_0$ is $g_{v1}(t)$, for $v_1$ and $v_2$ is $g_{v2}(t)$. From the theory of adiabatic state transfer, there should be $\frac{g_{v1}(t)}{g_{v2}(t)}=-\frac{\langle\hat{v}_2(t)\rangle}{\langle\hat{a}_0(t)\rangle}$, where $|\langle\hat{v}_2(t)\rangle|^2$ ($|\langle\hat{a}_0(t)\rangle|^2$) is the probability that the system is in state $|00...01\rangle_{a_0a_1...v_1v_2}$ ($|10...00\rangle_{a_0a_1...v_1v_2}$) at time $t$. In the input process, at time $t=0$, the virtual cavity $v_2$ is in Fock state $|1\rangle$, and the modes $a_j$ $(j=0,1,2,...,n)$, $b_m$ and $v_1$ are all in the empty states. At any time $t\in[0,T]$, there are $\langle\hat{a}_{0}(t)\rangle=\frac{-f(T-t)}{\sqrt{\kappa_0}}$ and $\langle\hat{v}_2(t)\rangle=\sqrt{\int_{t}^{T}|f(T-t)|^{2} dt}$, so that, there should be $\frac{g_{v1}(t)}{g_{v2}(t)}=\sqrt{\kappa_0}\frac{\sqrt{\int_{t}^{T}|f(T-t)|^{2} dt}}{f(T-t)}$.

The linearized interaction Hamiltonian of the system with virtual cavities is ($\hbar=1$) 

\begin{equation}
\hat{H}_{tot} = \hat{H}_I+(g_{v1}\hat{a}_0^{\dagger}\hat{v}_1+g_{v2}\hat{v}_{1}^{\dagger}\hat{v}_{2}+H.c.)
\end{equation}

With the mechanical resonator to be cooled to vacuum initially, the density matrix of the total system evolves according to the Lindblad master equation,
\begin{equation}
\frac{d\rho_{tot}}{dt}=-i[\hat{H}_{tot},\rho]+\sum_{j=1}^{n+1}(\hat{L}_{j}\rho\hat{L}_{j}^{\dagger}-\frac{1}{2}\{\hat{L}_{j}^{\dagger}\hat{L}_{j},\rho\}),
\end{equation}
where for $j = 1,2,...,n$, $\hat{L}_j = \sqrt{\kappa_{j}}\hat{a}_{j}$, and for $j=n+1$, $\hat{L}_{n+1} = \sqrt{\gamma_{m}}\hat{b}_{m}$.

The density matrix for the cavities $a_1$ to $a_n$ can be obtained by tracing out the modes $b_m$, $a_0$,$v_1$ and $v_2$ as $\rho_{w}=tr_{b_m,a_0,v_1,v_2}(\rho_{tot})$. Assume that, the density matrix for the ideal W state is $\rho_{w_{0}}$, the fidelity for the input process is $F = tr(\rho_{w_{0}}\rho_{w})$.

\begin{figure}[h]
\centering{\includegraphics[width=85mm]{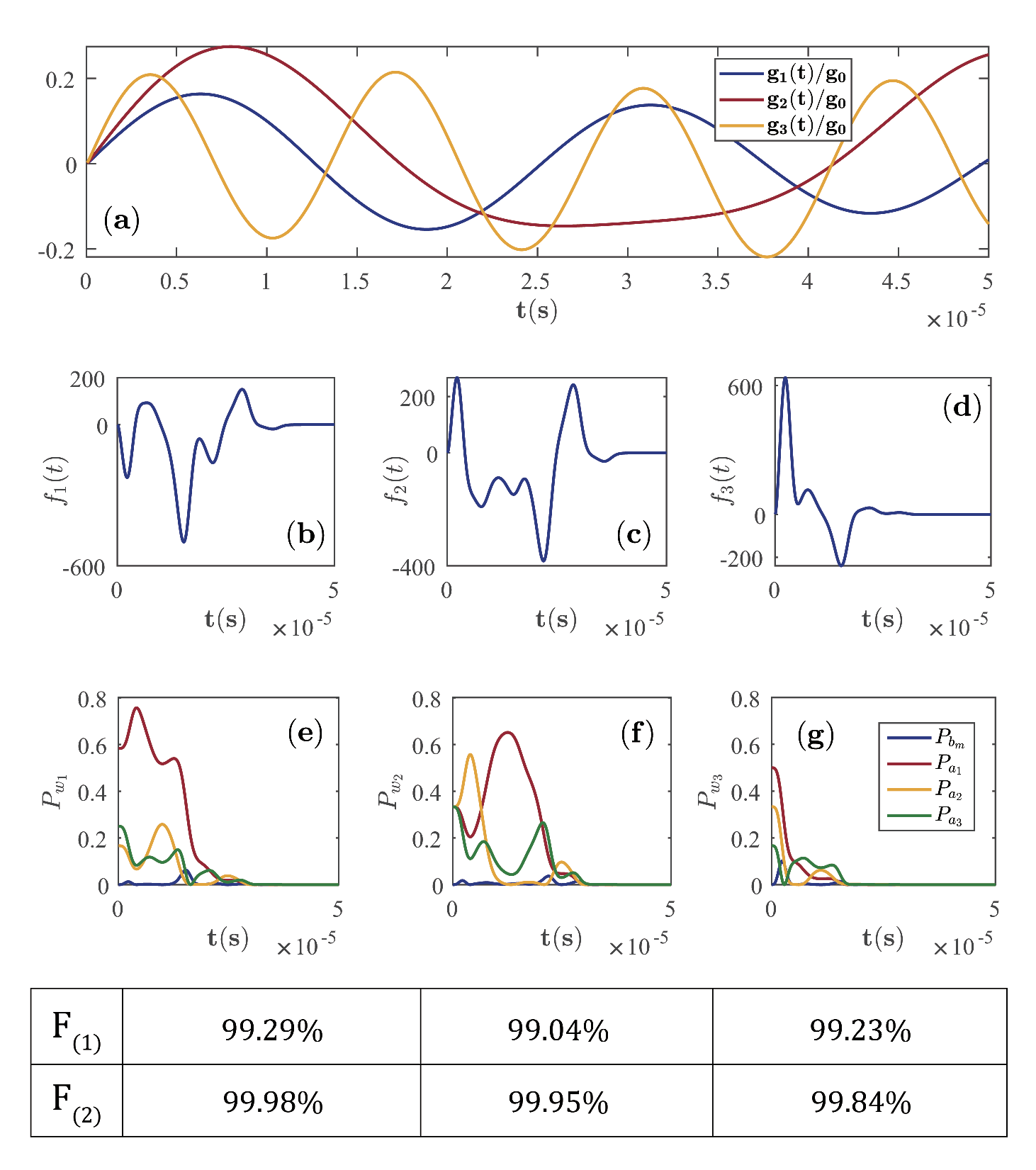}}
\caption{Numerical simulation results of an example for the processes of transmission and generation of entangled W states with $n=3$. In the beginning for the generation process, the states of the system are in $|W_1\rangle$, $|W_2\rangle $ or $|W_3\rangle $. (a) The optimal time-dependent couplings. (b)-(d) The functions of the single photon pulse shape with the initial states of $|W_1\rangle$, $|W_2\rangle $ and $|W_3\rangle$. (e)-(g) show the populations of the output process for the initial state $W_1$ to $W_3$. $P_d$ is the probability of the mode $\hat{d}$ ($d = b_m, a_1, a_2, a_3$) being in the Fock state $|1\rangle$. $F_{(1)}$ ($F_{(2)}$) is the fidelity of the whole process including the output and receiving processes with (without) dampings. For $F_{(1)}$, there are $\kappa_1/2\pi,\kappa_2/2\pi,\kappa_3/2\pi,\gamma_m/2\pi=10^2 \,\rm{Hz}, 10^2\,\rm{Hz}, 10^2\,\rm{Hz}, 10^4\,\rm{Hz}$ and for $F_{(2)}$, all the dampings equal to $0$.}
\label{f2}
\end{figure}

\begin{figure}[h]
\centering{\includegraphics[width=90mm]{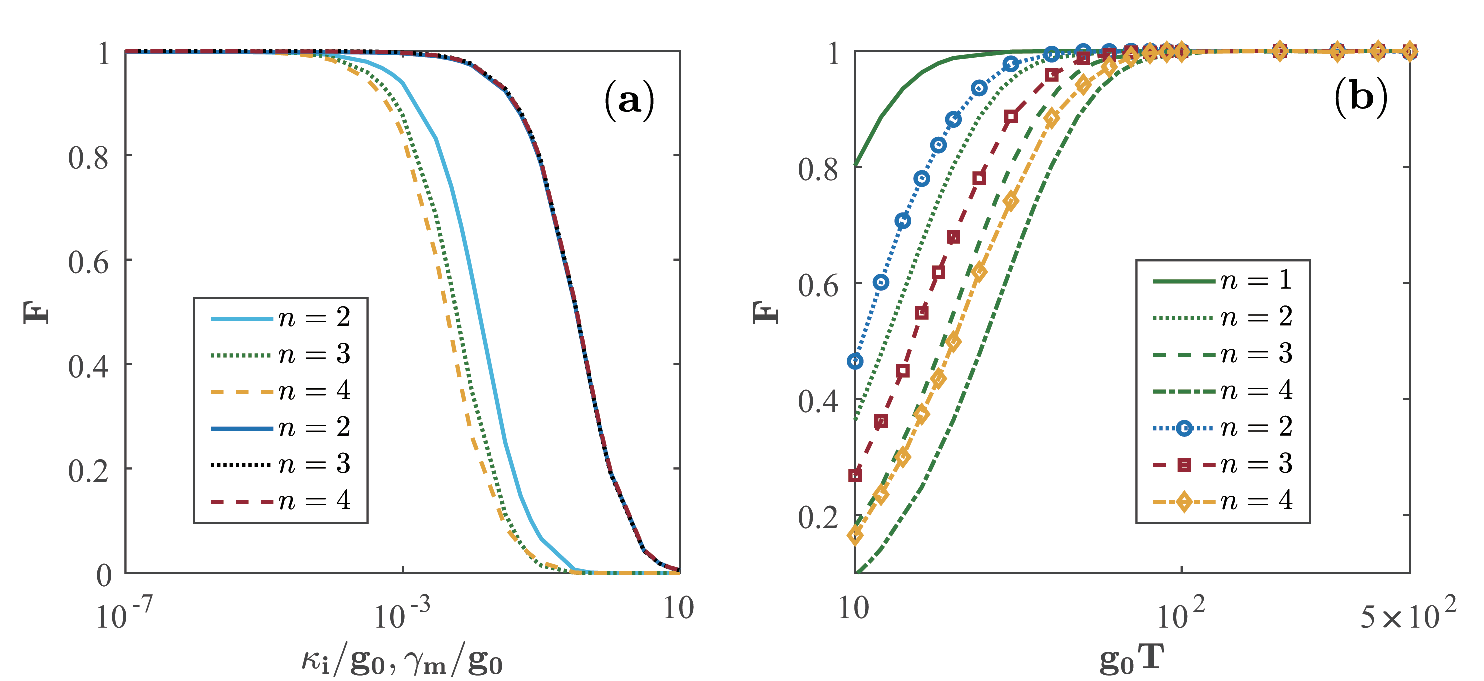}}
\caption{Fidelity for the generation process. The damping rate $\kappa_0=10g_0$. (a) Fidelity versus $\kappa_i/g_0$ (solid light blue, dotted light green and dashed yellow, $i=1,2,...,n$) with $\gamma_m=0$ and $g_0T=10^{2}$, and fidelity versus $\gamma_m/g_0$ (solid blue, dotted black, and dashed red) with $\kappa_i=0$ and $g_0T=10^{2}$. (b) Fidelity versus $g_0T$ with $\kappa_i=0$ and $\gamma_m=0$. The solid green line is for the system with only one microwave cavity. The dotted, dashed and dash dotted green lines are for the trivial method with reference to the optimized result of $n=1$. The dotted blue, dashed red and dash dotted yellow lines are the optimized results with the nontrivial method we proposed in this letter.}
\label{fidelity}
\end{figure}

\bigskip

\section{Numerical example}\label{s5}
A numerical example for adiabatic entangled W state transmission and generation is shown in Fig. \ref{f2}. In this system, the number of the microwave cavities is $3$ and the entangled W state is prepared in these cavities. The couplings $g_i$ ($i=1,2,3$) are time-dependent and optimized using CRAB optimization method \cite{doria2011optimal}. In the optimization process, $g_0$ is constant with a value of $10^7$ and $g_i(t)=\frac{1}{m}\sum_{k=1}^{m}A_{k}^{(i)}\sin{[2\pi k(1+r_{k})t/T]}$, where $m\in N_{+}$, $r_k\in[0,1]$ are random numbers and $A_k^{(i)}$ are optimizable parameters. The sine function can keep the initial values (at $ t=0$) of $g_i$ to be zero, and $g_i$ can also be expanded into other functions to meet different needs. The damping rate $\kappa_0/2\pi$ of the optical cavity is $0.1\, \rm GHz$, the mechanical resonator occurs at  $\omega_m/2\pi=1 \,\rm GHz$, and the phonon quantum state at GHz frequencies have been demonstrated in several experiments \cite{bienfait2019phonon,arrangoiz2019resolving,chu2018creation,satzinger2018quantum,chu2017quantum,o2010quantum}. The resolved-band condition of $\omega_m\gg\kappa_0$ is fulfilled, and in the adiabatic transmission and generation processes, the mechanical mode is a dark mode, the damping rate of which has less effect on the whole processes than the cavity dampings. We have calculated the fidelities for the generation process with two different sets of damping rate values, the first set is  $\kappa_1/2\pi,\kappa_2/2\pi,\kappa_3 /2\pi,\gamma_m/2\pi=10^2\, \rm{Hz}, 10^2 \,\rm{Hz}, 10^2 \,\rm{Hz}, 10^4 \,\rm{Hz}$ with internal quality factor of the cavities which can be realized in experiment \cite{megrant2012planar}, and the second is the ideal case for all of the dampings being set to zero.

In the beginning of the transmission process, the system is in an ideal W state. Here, we choose three different initial states as $|W_1\rangle = \frac{\sqrt{7}}{2\sqrt{3}}|100\rangle_{a_1a_2a_3}-\frac{1}{\sqrt{6}}|010\rangle_{a_1a_2a_3}-\frac{1}{2}|001\rangle_{a_1a_2a_3}$, $|W_2\rangle = \frac{1}{\sqrt{3}}|100\rangle_{a_1a_2a_3}-\frac{1}{\sqrt{3}}|010\rangle_{a_1a_2a_3}+\frac{1}{\sqrt{3}}|001\rangle_{a_1a_2a_3}$, and $|W_3\rangle = \frac{1}{\sqrt{2}}|100\rangle_{a_1a_2a_3}+\frac{1}{\sqrt{3}}|010\rangle_{a_1a_2a_3}+\frac{1}{\sqrt{6}}|001\rangle_{a_1a_2a_3}$. Based on the theoretical results of Section \ref{3b}, we optimized the time-dependent coupling strength with the output process, which are shown in Fig. \ref{f2}(a). With the optimal couplings, the arbitrary W state can be mapped into a single photon with specific wave function and transferred into another system, as shown in Fig. \ref{f1}(a), the single photon with certain wave function is transferred from system A to B. For the receiver, system B, they only need to know the functions of the time-depended coupling strength $g_i(t)$ which the system A used in the output process and need not know the wave function of the single photon, and then apply $g_i(T-t)$ to receive the single photon and generate a W state which is the same with the previous W state of system A.
 
 Fig. \ref{f2}(b)-(d) show the ensemble averaged pulse shape  $f_1(t)$, $f_2(t)$ and $f_3(t)$ of the single photons leaving from the optical cavity $a_0$ with the initial state of $|W_1\rangle$, $|W_2\rangle$ and $|W_3\rangle$, respectively. The ensemble averaged pulse shape can be expressed as $f(t)=-\sqrt{\kappa_0}\langle \hat{a}_0(t)\rangle$. In Fig. \ref{f2}(e)-(g), $P_{d}$ shows the numerical results of the probabilities where there is one population with mode $\hat{d}$ ($d = a_1, a_2, a_3, b_m$) in the output process for the states $W_1$, $W_2$ and $W_3$. For the input processes, we define the fidelity as $F_{(i)}=tr(\rho_{w_0}\rho_{w})$, $\rho_{w_0}$ is the original density matrix for the W state before transmission and $\rho_{w}$ is the final state after the receiving process. $F_{(1)}$ represents the process with the dampings  $\kappa_1/2\pi,\kappa_2/2\pi,\kappa_3/2\pi,\gamma_m/2\pi=10^2 \,\rm{Hz}, 10^2\,\rm{Hz}, 10^2\,\rm{Hz}, 10^4\,\rm{Hz}$ and $F_{(2)}$ represents the ideal process with dampings equal to $0$. The fidelities for transmitting the states $W_1$, $W_2$ and $W_3$ are $F_1=99.29\% (99.98\%)$,  $F_2=99.04\% (99.95\%)$ and $F_3 = 99.23\% (99.84\%)$ with (without) dampings.

\bigskip
\section{Fidelity for the generation process}\label{s6}
Since the generation process of W states is the time reversal of the transmission process, in this section, we talk about the effects of the damping and total time duration on the average fidelity for the generation process. Define the fidelity $F=tr(\rho_{w_0}\rho_{w})$ for the system with $n$ microwave cavities, $\rho_{w_0}$ is the density matrix for the ideal W state before the output process. $\rho_w$ is the density matrix of the final state after the input process for generating the target W state with a group of optimized time-dependent couplings.
In the entangled state transmission and generation processes, when $\kappa_i\ll\kappa_0,g_j$ ($i=1,2,...,n$, $j=0,1,2,...,n$), the whole process can be proceeded with high fidelity in a short time duration $T$, where $T\ll\frac{1}{\kappa_i}$. In the adiabatic transmission and generation processes, the mechanical mode is a dark mode, and high fidelity can be archived with the mechanical damping rate $\gamma_m<10^{-2}g_0$. However, when $\kappa_i$ and $\gamma_m$ become larger, the fidelity will reduce significantly, the numerical simulation results are shown in Fig. \ref{fidelity}(a).

We again consider the fidelity for the systems with different number of microwave cavities. With the same values of $g_0$, $\kappa_j$ ($j=0,1,2,...,n$) and $\gamma_m$,  in the transmission and generation process, using the trivial method, we can estimate that, to reach the same value of fidelity, it takes longer for the system with more microwave cavities as $T_n=nT_1$, where $T_n$ ($T_1$) is the time length for the system with $n$ (one) microwave cavities. Fig. \ref{fidelity}(b) shows the fidelity over the time length $T$ with the number of the microwave cavities $n=1,2,3,4$. The green lines are for the trivial method, which is mentioned detailedly at the end of Section \ref{s3}, and the blue, red and yellow lines are for the nontrivial method we proposed in this letter. With both of the two methods, for the system with more microwave cavities, it will take longer to get the same value of the fidelity, but the nontrivial method shows better performance than the trivial method.

\bigskip
\section{Mechanical Noise analysis}\label{s7}
In our scheme, we hope the system to be working in the mechanical dark state, which has been demonstrated to be immune to the mechanical noise \cite{tian2012adiabatic}. However, in order to balance the dissipation of the system and the adiabaticity of the evolving, there will be small probability for the mechanical mode to be excited. In the above calculations, we assume that, the mechanical resonator can be cooled to ground state initially, and which is not hard to realize in our scheme with the coupling between the mechanical resonator and the cavity $a_0$.

\begin{figure}[h!]
\centering{\includegraphics[width=75mm]{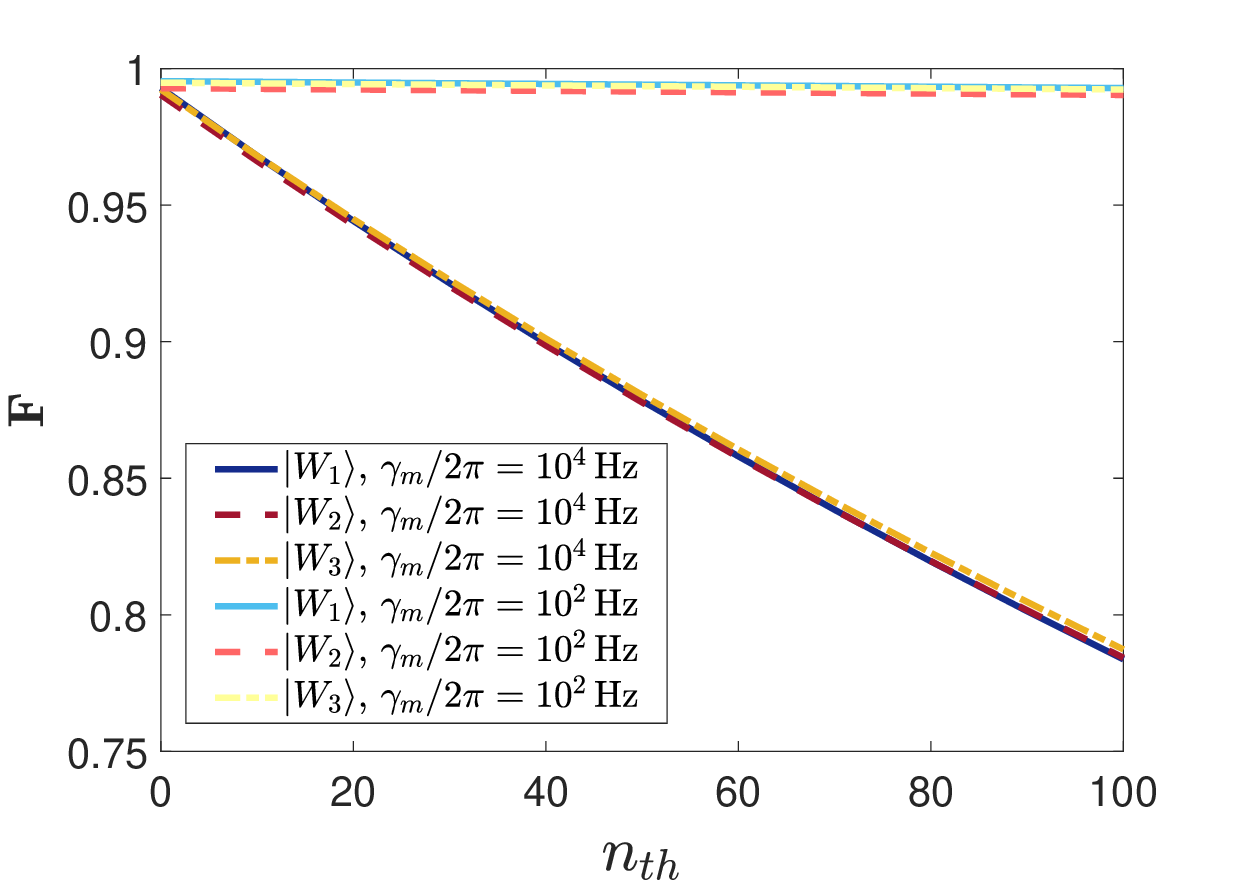}}
\caption{The fidelity of the generation process with mechanical noise for three different entangled states $|W_1\rangle$, $|W_2\rangle$ and $|W_3\rangle$. $n_{th}$ is the average phonon number. The mechanical decay rate is $\gamma_m/2\pi=10^4\,\rm Hz$ for solid blue, dashed red and dash-dotted yellow lines and $\gamma_m/2\pi=10^2\,\rm Hz$ for light solid blue, light dashed red and light dash-dotted yellow lines. The values of the other parameters are the same with the processes of $F_{(1)}$ in Fig. \ref{f2}.}
\label{f4}
\end{figure}

And in this section, we would like to study the effect of the mechanical noise when we do not cool the mechanical mode at the beginning of the whole process. The Lindblad master equation of the whole system with mechanical noise is,
\begin{eqnarray}
\frac{d\rho_{tot}}{dt}&=&-i[\hat{H}_{tot},\rho]+\sum_{j=1}^{n}(\hat{L}_{j}\rho\hat{L}_{j}^{\dagger}-\frac{1}{2}\{\hat{L}_{j}^{\dagger}\hat{L}_{j},\rho\})\nonumber\\
&+&(n_{th}+1)(\hat{L}_{n+1}\rho\hat{L}_{n+1}^{\dagger}-\frac{1}{2}\{\hat{L}_{n+1}^{\dagger}\hat{L}_{n+1},\rho\})\nonumber\\
&+&n_{th}(\hat{L}_{n+1}^{\dagger}\rho\hat{L}_{n+1}-\frac{1}{2}\{\hat{L}_{n+1}\hat{L}_{n+1}^{\dagger},\rho\}),
\end{eqnarray}
where $\hat{L}_j = \sqrt{\kappa_{j}}\hat{a}_{j}$ for the optical cavities $a_j$ ($j=1,2,...,n$), $\hat{L}_{n+1} = \sqrt{\gamma_{m}}\hat{b}_{m}$ for the mechanical resonator $b_m$ and $n_{th}$ is the thermal excitation number of the mechanical mode. Because the output is the inverse process of the input, we only study the input (generation) process in this section. In Fig. \ref{f4}, we plot the fidelities of the generation process for the states $|W_1\rangle$, $|W_2\rangle$ and $|W_3\rangle$ as the mechanical noise increases. When the decay rate of the mechanical mode is relatively large, with $\gamma_m/2\pi = 10^4 \,\rm Hz$, The impact of the noise is significant, and when the decay rate $\gamma_m/2\pi$ reduces to $10^2 \,\rm Hz$, the mechanical noise has negligible effect. For different target $W$ states with the same systematic parameters, the fidelities are almost the same, which means that, the impact from the target states are limited.

Then, considering the mechanical noise, if we want to enhance the fidelity of the whole process, we can pay attention to the following aspects. First is to enhance the quality factor of the mechanical resonator to decrease the value of the damping rate $\gamma_m$. Second, we can choose the mechanical mode with higher frequencies, which will be in the states with less number of excited thermal phonons at the same temperature. And the third one is to do the cooling with the mechanical mode at the beginning of the whole process.

\bigskip
\section{Conclusions}
In this letter, we have proposed a universal and nontrivial method for transmission and generation of the arbitrary entangled W states in multiple microwave cavities via an optomechanical interface. We have demonstrated that, the arbitrary entangled states can be mapped to the pulse shape of single photons and transmitted out of the system, which can be applied for the quantum information transfer. And the entangled W states can also be generated with the single incident photons with certain pulse shapes. In the generation process, the receiver need not know the wave function of the single photon or what the W state is, by using the specific time-dependent couplings, the receiver can receive a single photon with arbitrary wave function and produce a W state which is the same as the previous one in the sender's system with high fidelity. In the transmission and generation process, the wave functions (or ensemble averaged pulse shapes) of the single photons just act as a label for different W states, and using a certain group of optimal time-dependent couplings, the labels can be recognized and remapped to the W states in new systems. In our method, all of the microwave cavities are equivalent in time sequence, and it shows better performance in fidelity than the trivial method. Apart from the above results, there are still some open questions. The first one is how can we generate a state $W_b$ from an incident single photon which is produced by a different state $W_a$? And the second one is how to transmit the entangled quantum states with more than one excitations? Solving these open problems will also be helpful with the quantum information transfer between different quantum systems as well as the distributed quantum computing or quantum communication.

\bibliographystyle{apsrev4-1}
\bibliography{main}

\newpage 	
\renewcommand{\thefigure}{M\arabic{figure}}	
\setcounter{figure}{0}

% \newpage   
% \clearpage 
% \medskip

\newpage   
\clearpage 
\appendix 
\onecolumngrid

\section{Quantum jump and conditional Hamiltonian}

The non-Hermitian Hamiltonian of the system is

\begin{equation}
\hat{H}_{c}=-\sum_{j=0}^{n}\frac{i\kappa_{j}}{2}\hat{a}_j^{\dagger}\hat{a}_j-\frac{i\gamma_{m}}{2}\hat{b}_{m}^{\dagger}\hat{b}_{m}+H.c.
\end{equation}
where $j=0,1,2,...,n$, and we assume that $\hbar=1$, $H$ is the Hamiltonian for the closed system.

The probabilities of an emission from the system at time $t$ and during the time duration $\Delta t$ are

\begin{equation}
\Delta P_{j} = \kappa_{j} \Delta t \langle \psi(t)| \hat{a}_{j}^{\dagger}\hat{a}_{j} |\psi(t)\rangle,
\end{equation}

\begin{equation}
\Delta P_{m} = \gamma_{m} \Delta t \langle \psi(t)| \hat{b}_{m}^{\dagger}\hat{b}_{m} |\psi(t)\rangle,
\end{equation}
where $j=0,1,2,...,n$. So, the total probability for the quantum jump happening is $\Delta P_s = \sum_{j=0}^{n}\Delta P_{j}+\Delta P_{m}$.

And in fact, in the adiabatic process, the mechanical mode is in dark state, the probability of quantum happens for the mechanical mode will be far less than $1$, which can be ignored in the experiment.

If there is an emission from the cavity $a_i$, the system jumps to the renormalized state

\begin{equation}
\frac{\hat{a}_i|\psi(t)\rangle}{\sqrt{\langle\psi(t)|\hat{a}^{\dagger}_{i}\hat{a}_{i}|\psi(t)\rangle}}
\end{equation}

If there is an emission from the mechanical mode $b_m$, the system jumps to the renormalized state

\begin{equation}
\frac{\hat{b}_m|\psi(t)\rangle}{\sqrt{\langle\psi(t)|\hat{b}^{\dagger}_{m}\hat{b}_{m}|\psi(t)\rangle}}
\end{equation}

If there is no emission, the system will evolve depending on the non-Hermitian Hamiltonian as
\begin{eqnarray}
&&\frac{\exp{(-iH_c\Delta t)}|\psi(t)\rangle}{\sqrt{\langle\psi(t)|\exp{(-iH_c\Delta t)}^{\dagger}\exp{(-iH_c\Delta t)}|\psi(t)\rangle}}\nonumber\\
&\approx&\frac{(1-iH_c\Delta t)|\psi(t)\rangle}{\sqrt{\langle\psi(t)|(1-iH_c\Delta t)^{\dagger}(1-iH_c\Delta t)|\psi(t)\rangle}}\nonumber\\
&\approx&\frac{(1-\Delta t\sum_{j=0}^{n}\frac{\kappa_{j}}{2}\hat{a}_j^{\dagger}\hat{a}_j-\Delta t\frac{\gamma_{m}}{2}\hat{b}_{m}^{\dagger}\hat{b}_{m}-i\Delta t H)|\psi(t)\rangle}{\sqrt{\langle\psi(t)|(1-iH_c\Delta t)^{\dagger}(1-iH_c\Delta t)|\psi(t)\rangle}}\nonumber\\
&\approx&\frac{(1-\Delta t\sum_{j=0}^{n}\frac{\kappa_{j}}{2}\hat{a}_j^{\dagger}\hat{a}_j-\Delta t\frac{\gamma_{m}}{2}\hat{b}_{m}^{\dagger}\hat{b}_{m}-i\Delta t H)|\psi(t)\rangle}{\sqrt{1-\Delta t \sum_{j=0}^{n}\langle\psi(t)|\kappa_{j}\hat{a}_j^{\dagger}\hat{a}_j|\psi(t)\rangle-\Delta t \gamma_{m}\langle\psi(t)|\hat{b}^{\dagger}_{m}\hat{b}_m|\psi(t)\rangle}}\nonumber\\
&=&\frac{(1-\Delta t\sum_{j=0}^{n}\frac{\kappa_{j}}{2}\hat{a}_j^{\dagger}\hat{a}_j-\Delta t\frac{\gamma_{m}}{2}\hat{b}_{m}^{\dagger}\hat{b}_{m}-i\Delta t H)|\psi(t)\rangle}{\sqrt{1-\Delta P_s}}
\end{eqnarray}
where we have got rid of the high-order terms of $\Delta t$.

Then, after the time duration $\Delta t$, the density matrix will become

\begin{eqnarray}
&&|\psi(t+\Delta t)\rangle\langle\psi(t+\Delta t)|\nonumber\\
&=&\sum_{j=0}^{n}\Delta P_j \frac{\hat{a}_j|\psi(t)\rangle \langle\psi(t)|\hat{a}_{j}^{\dagger}}{\langle\psi(t)|\hat{a}_{j}^{\dagger}\hat{a}_{j}|\psi(t)\rangle}\
+\Delta P_m \frac{\hat{b}_m|\psi(t)\rangle\langle\psi(t)|\hat{b}_{m}^{\dagger}}{\langle\psi(t)|\hat{b}^{\dagger}_{m}\hat{b}_{m}|\psi(t)\rangle}\\
&+&(1-\Delta P_s) \frac{(1-\Delta t\sum_{j=0}^{n}\frac{\kappa_{j}}{2}\hat{a}_j^{\dagger}\hat{a}_j-\Delta t\frac{\gamma_{m}}{2}\hat{b}_{m}^{\dagger}\hat{b}_{m}-i\Delta t H)|\psi(t)\rangle \langle\psi(t)|(1-\Delta t\sum_{j=0}^{n}\frac{\kappa_{j}}{2}\hat{a}_j^{\dagger}\hat{a}_j-\Delta t\frac{\gamma_{m}}{2}\hat{b}_{m}^{\dagger}\hat{b}_{m}+i\Delta t H)}{1-\Delta P_s}\nonumber\\
&\approx& \Delta t \sum_{j=0}^{n}\kappa_{j} \hat{a}_j|\psi(t)\rangle \langle\psi(t)|\hat{a}_{j}^{\dagger} +\Delta t \gamma_{m} \hat{b}_m|\psi(t)\rangle\langle\psi(t)|\hat{b}_{m}^{\dagger} \\
&+& |\psi(t)\rangle\langle\psi(t)| - i\Delta t H |\psi(t)\rangle\langle\psi(t)| + i\Delta t |\psi(t)\rangle\langle\psi(t)| H - \Delta t\sum_{j=0}^{n}\frac{\kappa_{j}}{2}\hat{a}_j^{\dagger}\hat{a}_j|\psi(t)\rangle \langle\psi(t)| - \Delta t |\psi(t)\rangle \langle\psi(t)|\sum_{j=0}^{n}\frac{\kappa_{j}}{2}\hat{a}_j^{\dagger}\hat{a}_j \nonumber\\
&-& \Delta t \frac{\gamma_{m}}{{2}}\hat{b}_{m}^{\dagger}\hat{b}_{m} |\psi(t)\rangle\langle\psi(t)| - \Delta t \frac{\gamma_{m}}{{2}} |\psi(t)\rangle\langle\psi(t)| \hat{b}_{m}^{\dagger}\hat{b}_{m}\nonumber
\end{eqnarray}

The $(7)$ and $(8)$ come from the quantum jump, and the master equation can be written as

\begin{eqnarray}
\frac{\Delta\rho}{\Delta t}&=& \sum_{j=0}^{n}\kappa_{j} \hat{a}_j\rho\hat{a}_{j}^{\dagger} + \gamma_{m} \hat{b}_m\rho\hat{b}_{m}^{\dagger} \\
&+& - i H \rho + i \rho H - \sum_{j=0}^{n}\frac{\kappa_{j}}{2}\hat{a}_j^{\dagger}\hat{a}_j\rho - \sum_{j=0}^{n}\frac{\kappa_{j}}{2} \rho \hat{a}_j^{\dagger}\hat{a}_j \nonumber\\
&-&  \frac{\gamma_{m}}{{2}}\hat{b}_{m}^{\dagger}\hat{b}_{m} \rho - \frac{\gamma_{m}}{{2}} \rho \hat{b}_{m}^{\dagger}\hat{b}_{m}\nonumber\\
&=&-i[H,\rho]+\sum_{j=0}^{n}\frac{\kappa_j}{2}(2\hat{a}_{j}\rho\hat{a}^{\dagger}_{j}-\hat{a}^{\dagger}_{j}\hat{a}_{j}\rho-\rho\hat{a}^{\dagger}_{j}\hat{a}_{j})+\frac{\gamma_m}{2}(2\hat{b}_{m}\rho\hat{b}^{\dagger}_{m}-\hat{b}^{\dagger}_{m}\hat{b}_{m}\rho-\rho\hat{b}^{\dagger}_{m}\hat{b}_{m})
\end{eqnarray}

In the basis of the bare state, we can calculate the master equation with the rotating wave approximation, then we can replace $H$ with the linearized interaction Hamiltonian $\hat{H}_I$, where $\hat{H}_I = \sum_{j=0}^{n}g_{j}\left(\hat{a}_{j}^{\dagger}\hat{b}_{m}+\hat{b}_{m}^{\dagger}\hat{a}_{j}\right)$. So that, if we eliminate the processes in which the quantum jump happened by applying the post selection, the non-Hermitian conditional Hamiltonian can be used to describe the evolution of the system as:

\begin{eqnarray}
\frac{\Delta\tilde{\rho}}{\Delta t}=-i[\hat{H}_c,\tilde{\rho}]
\end{eqnarray}

Where $\hat{H}_c=-\sum_{j=0}^{n}\frac{i\kappa_{j}}{2}\hat{a}_j^{\dagger}\hat{a}_j-\frac{i\gamma_{m}}{2}\hat{b}_{m}^{\dagger}\hat{b}_{m}+\hat{H}_I$. In fact, if only considering the coherent decay, the process is a coherent nonunitary evolution, it can also be described by the Schr{\"o}dinger equation with the conditional Hamiltonian.

\bigskip
\section{The conditional Hamiltonian in the adiabatic evolution}

Under the condition of $g_i\gg\kappa_i$ $(i = 1, 2, ..., n)$, for simplicity, we first neglect $\kappa_i$ and only consider the damping of $\kappa_0$ and $\gamma_m$. In the basis of $|100...0\rangle _{a_{0}a_{1}...a_{n}b_{m}}$, $|010...0\rangle _{a_{0}a_{1}...a_{n}b_{m}}$,..., $|000...1\rangle _{a_{0}a_{1}...a_{n}b_{m}}$, the conditional Hamiltonian is

\begin{equation}
\hat{H}_{c}=\left[\begin{array}{ccccc}
-\frac{i\kappa_{0}}{2} & 0 & \cdots & 0 & g_{0}\\
0 & 0 & \cdots & 0 & g_{1}\\
\vdots & \vdots & \vdots & \vdots & \vdots\\
0 & 0 & \cdots & 0 & g_{n}\\
g_{0} & g_{1} & \cdots & g_{n} & -\frac{i\gamma_{m}}{2}
\end{array}\right]
\label{hc1}
\end{equation}

If we transform the basis to $|\phi_{1}(t)\rangle $, $|\phi_{2}(t)\rangle $,...,$|\phi_{n+2}(t)\rangle $, where $|\phi_{1}(t)\rangle $ to $|\phi_{n}(t)\rangle $ are $n$ dark states and $|\phi_{n+1}(t)\rangle $ and $|\phi_{n+2}(t)\rangle $ are two bright states, the conditional Hamiltonian can be transformed as

\begin{equation}
\hat{\tilde{H}}_{c}^{[(n+2)\times(n+2)]}=\left[\begin{array}{ccccccc}
-\frac{i\kappa_0 g_{1}^{2}}{2s_{1}^{2}} & -\frac{i\kappa_0 g_{0}g_{1}g_{2}}{2s_{1}^{2}s_{2}} & -\frac{i\kappa_0 g_{0}g_{1}g_{3}}{2s_{1}s_{2}s_{3}} & \cdots & -\frac{i\kappa_0 g_{0}g_{1}g_{n}}{2s_{1}s_{n-1}s_{n}} & \tilde{H}_{c}^{(1,n+1)} & \tilde{H}_{c}^{(1,n+2)}\\
-\frac{i\kappa_0 g_{0}g_{1}g_{2}}{2s_{1}^{2}s_{2}} & -\frac{i\kappa_0 g_{0}^{2}g_{2}^{2}}{2s_{1}^{2}s_{2}^{2}} & -\frac{i\kappa_0 g_{0}^{2}g_{2}g_{3}}{2s_{1}s_{2}^{2}s_{3}} & \cdots & -\frac{i\kappa_0 g_{0}^{2}g_{2}g_{n}}{2s_{1}s_{2}s_{n-1}s_{n}} & \tilde{H}_{c}^{(2,n+1)} & \tilde{H}_{c}^{(2,n+2)}\\
-\frac{i\kappa_0 g_{0}g_{1}g_{3}}{2s_{1}s_{2}s_{3}} & -\frac{i\kappa_0 g_{0}^{2}g_{2}g_{3}}{2s_{1}s_{2}^{2}s_{3}} & -\frac{i\kappa_0 g_{0}^{2}g_{3}^{2}}{2s_{2}^{2}s_{3}^{2}} & \cdots & -\frac{i\kappa_0 g_{0}^{2}g_{3}g_{n}}{2s_{2}s_{3}s_{n-1}s_{n}} & \tilde{H}_{c}^{(3,n+1)} & \tilde{H}_{c}^{(3,n+2)}\\
\vdots & \vdots & \vdots & \vdots & \vdots & \vdots & \vdots\\
-\frac{i\kappa_0 g_{0}g_{1}g_{n-1}}{2s_{1}s_{n-2}s_{n-1}} & -\frac{i\kappa_0 g_{0}^{2}g_{2}g_{n-1}}{2s_{1}s_{2}s_{n-2}s_{n-1}} & -\frac{i\kappa_0 g_{0}^{2}g_{3}g_{n-1}}{2s_{2}s_{3}s_{n-2}s_{n-1}} & \cdots & -\frac{i\kappa_0 g_{0}^{2}g_{n-1}g_{n}}{2s_{n-2}s_{n-1}^{2}s_{n}} & \tilde{H}_{c}^{(n-1,n+1)} & \tilde{H}_{c}^{(n-1,n+2)}\\
-\frac{i\kappa_0 g_{0}g_{1}g_{n}}{2s_{1}s_{n-1}s_{n}} & -\frac{i\kappa_0 g_{0}^{2}g_{2}g_{n}}{2s_{1}s_{2}s_{n-1}s_{n}} & -\frac{i\kappa_0 g_{0}^{2}g_{3}g_{n}}{2s_{2}s_{3}s_{n-1}s_{n}} & \cdots & -\frac{i\kappa_0 g_{0}^{2}g_{n}^{2}}{2s_{n-1}^{2}s_{n}^{2}} & \tilde{H}_{c}^{(n,n+1)} & \tilde{H}_{c}^{(n,n+2)}\\
\tilde{H}_{c}^{(n+1,1)} & \tilde{H}_{c}^{(n+1,2)} & \tilde{H}_{c}^{(n+1,3)} & \cdots & \tilde{H}_{c}^{(n+1,n)} & \tilde{H}_{c}^{(n+1,n+1)} & \tilde{H}_{c}^{(n+1,n+2)}\\
\tilde{H}_{c}^{(n+2,1)} & \tilde{H}_{c}^{(n+2,2)} & \tilde{H}_{c}^{(n+2,3)} & \cdots & \tilde{H}_{c}^{(n+2,n)} & \tilde{H}_{c}^{(n+2,n+1)} & \tilde{H}_{c}^{(n+2,n+2)}
\end{array}\right]
\end{equation}
where $(\hat{\tilde{H}}_c^{[(n+2)\times(n+2)]})_{ij}=\langle\phi_i|\hat{H}_c|\phi_j\rangle$.

In the adiabatic process, initially, if the system is in dark
state, there will not be excitations with the bright states, so that,
the coherent decay evolution of the system will only depend on the first $n$ rows
and $n$ columns of the conditional Hamiltonian. We extract the first $n$ rows and $n$ columns
of the matrix of $\hat{\tilde{H}}_{c}$ as

\begin{eqnarray}
\hat{\tilde{H}}_{c}^{(n\times n)} & = & \left[\begin{array}{ccccc}
-\frac{i\kappa_0 g_{1}^{2}}{2s_{1}^{2}} & -\frac{i\kappa_0 g_{0}g_{1}g_{2}}{2s_{1}^{2}s_{2}} & -\frac{i\kappa_0 g_{0}g_{1}g_{3}}{2s_{1}s_{2}s_{3}} & \cdots & -\frac{i\kappa_0 g_{0}g_{1}g_{n}}{2s_{1}s_{n-1}s_{n}}\\
-\frac{i\kappa_0 g_{0}g_{1}g_{2}}{2s_{1}^{2}s_{2}} & -\frac{i\kappa_0 g_{0}^{2}g_{2}^{2}}{2s_{1}^{2}s_{2}^{2}} & -\frac{i\kappa_0 g_{0}^{2}g_{2}g_{3}}{2s_{1}s_{2}^{2}s_{3}} & \cdots & -\frac{i\kappa_0 g_{0}^{2}g_{2}g_{n}}{2s_{1}s_{2}s_{n-1}s_{n}}\\
-\frac{i\kappa_0 g_{0}g_{1}g_{3}}{2s_{1}s_{2}s_{3}} & -\frac{i\kappa_0 g_{0}^{2}g_{2}g_{3}}{2s_{1}s_{2}^{2}s_{3}} & -\frac{i\kappa_0 g_{0}^{2}g_{3}^{2}}{2s_{2}^{2}s_{3}^{2}} & \cdots & -\frac{i\kappa_0 g_{0}^{2}g_{3}g_{n}}{2s_{2}s_{3}s_{n-1}s_{n}}\\
\vdots & \vdots & \vdots & \vdots & \vdots\\
-\frac{i\kappa_0 g_{0}g_{1}g_{n-1}}{2s_{1}s_{n-2}s_{n-1}} & -\frac{i\kappa_0 g_{0}^{2}g_{2}g_{n-1}}{2s_{1}s_{2}s_{n-2}s_{n-1}} & -\frac{i\kappa_0 g_{0}^{2}g_{3}g_{n-1}}{2s_{2}s_{3}s_{n-2}s_{n-1}} & \cdots & -\frac{i\kappa_0 g_{0}^{2}g_{n-1}g_{n}}{2s_{n-2}s_{n-1}^{2}s_{n}}\\
-\frac{i\kappa_0 g_{0}g_{1}g_{n}}{2s_{1}s_{n-1}s_{n}} & -\frac{i\kappa_0 g_{0}^{2}g_{2}g_{n}}{2s_{1}s_{2}s_{n-1}s_{n}} & -\frac{i\kappa_0 g_{0}^{2}g_{3}g_{n}}{2s_{2}s_{3}s_{n-1}s_{n}} & \cdots & -\frac{i\kappa_0 g_{0}^{2}g_{n}^{2}}{2s_{n-1}^{2}s_{n}^{2}}
\end{array}\right]\nonumber\\
 & = & -\frac{i\kappa_0}{2}M
\end{eqnarray}
where $M=(1-\frac{g_{0}^{2}}{s_{n}^{2}})|\phi_{0}\rangle \langle \phi_{0}|$ and $|\phi_{0}\rangle =\frac{1}{\sqrt{1-\frac{g_{0}^{2}}{s_{n}^{2}}}}[
\phi_{1}^{(1)},\phi_{2}^{(1)},\phi_{3}^{(1)},...,\phi_{n}^{(1)}]^{T}$, where $\phi_{i}^{(1)}$ ($i=1,2,...,n$) is the first value of $|\phi_{i}\rangle $.

\bigskip
\section{The characteristics of V matrix}

$V(t)$ matrix is defined as $V(t)=\frac{dU^{\dagger}(t)}{dt}U(t)$ and $\alpha(t)=U^{\dagger}(t)C(t)$, and there is

\begin{equation}
\frac{d\alpha(t)}{dt}=-[\frac{\kappa_0}{2}\Lambda(t)-V(t)]\alpha(t)
\end{equation}

Because $\frac{d}{dt}[U^{\dagger}(t)U(t)]=0$,
so we can get $\frac{dU^{\dagger}(t)}{dt}U(t)+U^{\dagger}(t)\frac{dU^{\dagger}(t)}{dt}=0$,
then there is $V(t)+V^{\dagger}(t)=0$. Since that, the matrix $V(t)$ is real, we have $V^{\dagger}(t)=V^{T}(t)$, so we can get $V_{ij}(t)+V_{ji}(t)=0$. If $\kappa_0=0$, there is

\begin{eqnarray}
\frac{d}{dt}\sum_{i=1}^{n}\alpha_{i}^{2}(t) & = & 2\sum_{i=1}^{n}\alpha_{i}(t)\frac{d\alpha_{i}\left(t\right)}{dt}\nonumber\\
 & = & 2\sum_{i=1}^{n}\alpha_{i}(t)[\sum_{j=1}^{n}V_{ij}(t)\alpha_{j}(t)]\nonumber\\
 & = & 2\sum_{i,j=1}^{n}V_{ij}(t)\alpha_{i}(t)\alpha_{j}(t)\nonumber\\
 & = & 0
\end{eqnarray}

So that, the function of $V(t)$ is only to redistribute of the populations in every $\varphi_{i}(t)$.

\end{document}